\documentclass[letterpaper,twocolumn,prl,showpacs]{revtex4-1}
\usepackage{graphicx,amsmath,amssymb}

\renewcommand{\[}{\begin{equation}}
\renewcommand{\]}{\end{equation}}
\def\bea{\begin{eqnarray}}
\def\eea{\end{eqnarray}}
\def\nn{\nonumber\\}
\newcommand{\equ}[1]{Eq.~(\ref{#1})}
\newcommand{\eqs}[2]{Eqs.~(\ref{#1}) and (\ref{#2})}

\renewcommand{\k}{{\bf k}}
\renewcommand{\r}{{\bf r}}
\renewcommand{\j}{{\bf j}}
\newcommand{\M}{{\bf M}}

\renewcommand{\P}{{\bf P}}
\newcommand{\CQ}{{\cal Q}}
\newcommand{\CP}{{\cal P}}
\def\MM{{\mathfrak M}}
\def\CC{{\mathfrak C}}
\def\beq{\begin{equation}}
\def\eeq{\end{equation}}
\def\bea{\begin{eqnarray}}
\def\eea{\end{eqnarray}}
\def\nn{\nonumber\\}
\newcommand{\intr}{\int \! d{\bf r} \;}

\renewcommand{\r}{{\bf r}}

\def\bra#1{\langle#1\vert}
\def\ket#1{\vert#1\rangle}

\def\me#1#2#3{\langle#1| \, #2 \, |#3\rangle}
\newcommand{\smu}{\sum_{\epsilon_n < \mu}}

\def\runtime{(\the\time)\qquad\the\month/\the\day/\the\year}
\def\today
 {\count10=\year\advance\count10 by -2000 \number\day--\ifcase
  \month \or Jan\or Feb\or Mar\or Apr\or May\or Jun\or
             Jul\or Aug\or Sep\or Oct\or Nov\or Dec\fi--\number\count10}

\def\hour{\count10=\time\count11=\count10
\divide\count10 by 60 \count12=\count10
\multiply\count12 by 60 \advance\count11 by -\count12\count12=0
\number\count10 :\ifnum\count11 < 10 \number\count12\fi\number\count11}

\begin{document}

\title{Orbital Magnetization as a Local Property}

\author{Raffaello Bianco$^1$ and Raffaele Resta$^{1,2}$}

\affiliation{$^1$Dipartimento di Fisica, Universit\`a di Trieste, Italy}
\affiliation{$^2$ Centre Europ\'een de Calcul Atomique et Mol\'eculaire (CECAM), \'Ecole Polytechnique F\'ed\'erale de Lausanne, Switzerland}

\begin{abstract}
The modern expressions for polarization $\P$ and orbital magnetization $\M$ are $\k$-space integrals. But a genuine bulk property should also be expressible in $\r$-space, as unambiguous function of the ground-state density matrix, ``nearsighted'' in insulators, independently of the boundary conditions---either periodic or open. While $\P$---owing to its ``quantum'' indeterminacy---is {\it not} a bulk property in this sense, $\M$ is. We provide its $\r$-space expression for any insulator, even with nonzero Chern invariant. Simulations on a model Hamiltonian validate our theory.
\end{abstract}

\date{\today}


\maketitle \bigskip\bigskip

The macroscopic polarization $\P$ and magnetization $\M$ are essential ingredients of the in-medium Maxwell equations, but microscopic understanding of $\P$ and of the orbital contribution to $\M$ was achieved only in recent times \cite{King93,Vanderbilt93,Xiao05,rap128,rap130,Souza08}. Their elementary definition for a finite sample is \bea  \P  &=&
\frac{{\bf d}}{V} = \frac{1}{V} \int d\r \; \r \rho^{\rm (micro)}(\r)\label{PP} \\ \M &=&
\frac{{\bf m}}{V} = \frac{1}{2cV} \int d\r \; \r \times \j^{\rm (micro)}(\r) . \label{PM} \eea  Here and in the following we indicate with $\M$ the orbital term only;  $\rho^{(\rm micro)}(\r)$ and $\j^{\rm (micro)}(\r)$ are the microscopic
charge and current densities, and $V$ is the sample volume. The previous expressions are clearly dominated by surface contributions, while instead phenomenologically $\P$ and $\M$ are bulk properties: from this viewpoint, the two properties appear as closely analogous.
The modern theories of polarization and magnetization (in their simplest formulations) address a crystalline system of independent electrons; therein, both $\P$ and $\M$ are expressed as a Brillouin-zone integral of Bloch-orbital matrix elements \cite{King93,Vanderbilt93,Xiao05,rap128,rap130,Souza08,rap_a12,rap_a30}; even the $\k$ space expressions for $\P$ and $\M$ share many analogies.
The modern theories are clearly based on periodic boundary conditions (PBCs): the sample has no boundary, and the properties are ``bulk'' by definition. 

In this Letter we aim instead at $\r$-space definitions, but where---at variance with \eqs{PP}{PM}---the choice of the boundary conditions becomes irrelevant in the limit of a large sample. For a system of independent electrons the ground state is uniquely determined by the one-particle density matrix, a.k.a. ground-state projector $\CP(\r,\r')$; it is a ``nearsighted'' \cite{Kohn96,rap132,rap_a31} operator, exponentially decreasing with $|\r - \r'|$ in insulators even when the Chern invariant is nonzero \cite{Thonhauser06}. Our aim is therefore to express $\P$ and $\M$ as local properties in $\r$ space, directly in terms of $\CP(\r,\r')$ in the bulk of a sample, {\it independently} of the boundary conditions. We show that such aim {\it cannot} be attained for $\P$, while we provide an explicit expression for $\M$, even for insulators with nonzero Chern invariant (``Chern insulators''). Tinkering with the boundaries may alter the value of $\P$, but not of $\M$: this finding is in agreement with a very recent work by Chen and Lee, based on completely different arguments \cite{Chen12}.

We validate our approach by means of simulations on a model Hamiltonian, performed on finite samples with {\it open} boundary conditions (OBCs). One outstanding virtue of our formula is that it converges to the bulk $\M$ value much faster than the elementary definition of \equ{PM}: see Fig. \ref{fig:m2} below; another virtue is that it could be applied with no major changes to disordered and/or macroscopically inhomogeneous systems.

The modern theory of polarization addresses the difference in polarization $\Delta \P$ between two states of the material that can be connected by an adiabatic switching process. This is clearly a bulk property, provided the system remains insulating at all times: $\Delta \P$ in fact coincides with the integrated current flow across the material, which in turn is easily expressed in terms of the evolution of $\CP(\r,\r')$ along the switching. But ``$\P$ itself'' is {\it not} a bulk property in the above sense: a basic tenet of the modern theory of polarization states that the bulk electron distribution determines $\P$ only {\it modulo a ``quantum''}, whose value depends on the boundary \cite{King93,Vanderbilt93}. Therefore it is impossible to evaluate $\P$ for a homogeneous sample knowing $\CP(\r,\r')$ in its bulk only: examples of systems with the same bulk and different $\P$ values are e.g. in Ref. \cite{rap136}. 

The modern theory of magnetization, instead, addresses ``$\M$ itself'' directly, and is not affected by any quantum indeterminacy. Therefore an expression for $\M$ in terms of the bulk density matrix (either PBCs or OBCs), ergo boundary-independent, is not ruled out. Here we are providing such  expression: in any macroscopically homogeneous sample $\M$ is the macroscopic average---defined as in electrostatics \cite{Jackson}---of a local function $\MM(\r)$, uniquely defined in terms of the density matrix in a neighborhood of $\r$. We draw attention to the fact that, in a polarized/magnetized solid, the charge and current densities
$\rho^{(\rm micro)}(\r)$ and $\j^{\rm (micro)}(\r)$ are well defined, while a ``dipolar density'' (either electric or magnetic) cannot be unambiguously defined \cite{Hirst97,nota1}; our $\MM(\r)$ plays indeed the role of a magnetic dipolar density, although only its macroscopic average bears a physical meaning.

The main concepts are more clearly formulated in the simple two-dimensional (2D) case: electrons in the $xy$ plane and magnetization $M$ along $z$. \equ{PM} reads,
for a 2D   macroscopic flake of independent electrons \bea M &=& - \frac{i e}{2 \hbar c A} \smu \me{\varphi_n}{\r \times [H,\r] \,
}{\varphi_n} \label{m2} \\ &=&  - \frac{i e}{2 \hbar c A} \smu ( \; \me{\varphi_n}{x H y}{\varphi_n} - \me{\varphi_n}{y H x
}{\varphi_n} \; ) \nonumber \eea where $A$ is the sample area, $H$ is the single-particle Hamiltonian, $\ket{\varphi_n}$ are the orbitals, and $\mu$ is the Fermi level; single occupancy is assumed (``spinless electrons''). 
\equ{m2} only applies to a system that remains gapped (as a whole) in the large-$A$ limit, and therefore does not apply, as such, to Chern insulators; more about this will be said below. \equ{m2} is a trace; since $M$ is real, \beq M = \mbox{Im } iM = \frac{e}{\hbar c A} \mbox{Im Tr }\{ \CP x H y \CP \}, \label{m3} \eeq where $\CP$ is the ground-state projector. In the following, we also need its complement $\CQ$, i.e. \beq \CP = \smu \ket{\varphi_n} \bra{\varphi_n}, \qquad \CQ = 1 - \CP . \eeq If we write $H = \CP H \CP + \CQ H \CQ$, it is rather straightforward to transform \equ{m3} into \beq M = \frac{e}{\hbar c A} \mbox{Im Tr } \{ \CP x \CQ H \CQ y  \CP -  \CQ x \CP H \CP y \CQ \} \label{m4} . \eeq A different derivation of the same expression is due to Souza and Vanderbilt \cite{Souza08}; they also show that \equ{m4} provides the link with the modern theory of magnetization. In fact the position operator $\r$ is ill  defined within PBCs \cite{rap100}, but becomes harmless and well defined within both OBCs and PBCs when ``sandwiched'' between a $\CP$ and a $\CQ$. It is enough to perform the thermodynamic limit in \equ{m4}, and then cast $\CP$ and $\CQ$ in terms of Bloch orbitals, in order to arrive at the $\k$-integral expression of the modern theory \cite{Xiao05,rap128,rap130} for normal insulators (Chern number $C=0$).

In order to get a local description we write \equ{m4} as \beq M = \frac{1}{A} \intr \MM_1(\r) , \label{local} \eeq \bea \MM_1(\r) &=& \frac{e}{\hbar c} \mbox{Im }  \me{\r}{\CP x \CQ H \CQ y \CP }{\r} \nn &-& \frac{e}{\hbar c} \mbox{Im }  \me{\r}{\CQ x \CP H \CP y \CQ }{\r}. \label{mpq}  \eea
There is a paramount difference between our starting \equ{m2} and \equ{local}: while the former integral, like \equ{PM}, is dominated by boundary contributions, the latter expression is ``bulk'' in the above defined sense. In order to evaluate $M$ for a macroscopically homogeneous region in the bulk of a sample, within either OBCs or PBCs, it is enough to take the macroscopic average of $\MM_1(\r)$ in that region.

\begin{figure}
\centering
\includegraphics[width=.7\columnwidth]{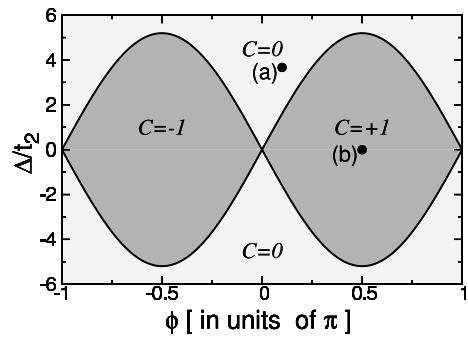}
\caption{Chern number $C$ of the bottom band of the Haldane model as a function of the parameters $\varphi$ and $\Delta/t_2$ ($t_1=1, t_2=1/3$). The subsequent discussion and figures concern the points (a) and (b) only}
\label{fig:diagram} \end{figure}

We demonstrate this key property of the local function $\MM_1(\r)$ by performing simulations on the Haldane model Hamiltonian \cite{Haldane88}; it is
comprised of a 2D    honeycomb lattice with two tight-binding sites per primitive
cell with site energies $\pm \Delta$, real first-neighbor hoppings $t_1$, and
complex second-neighbor hoppings $t_2e^{\pm i\varphi}$.  This model has been previously used in several simulations, providing invaluable insight into orbital magnetization \cite{rap128,rap130,rap135} as well as into nontrivial topological features of the electronic wavefunction \cite{Haldane88,rap135,Thonhauser06,Hao08,Coh09,rap146}. In the following, we invariably choose $t_1=1$, $t_2=1/3$. At half filling the system is insulating; it is either a normal insulator or a Chern insulator depending on the $\Delta$ and $\varphi$ values, according to the phase diagram shown in Fig. \ref{fig:diagram}.

\begin{figure}
\centering
\includegraphics[width=.9\columnwidth]{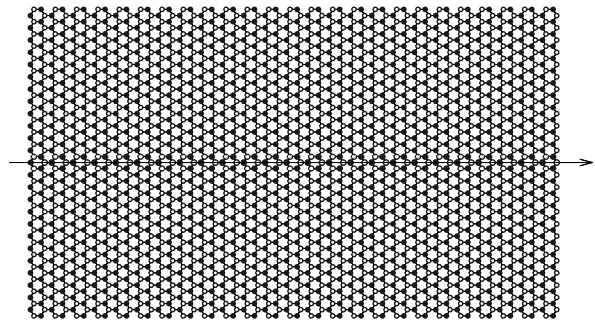}
\caption{A typical flake, with 2550  sites, showing the honeycomb lattice of the Haldane model \protect\cite{Haldane88}. The 50 sites on the horizontal line  will be used in all the subsequent one-dimensional plots. Black and grey circles indicate nonequivalent sites (with onsite energies $\pm \Delta$)}
\label{fig:flake} \end{figure}

\begin{figure}
\centering
\includegraphics[width=.8\columnwidth]{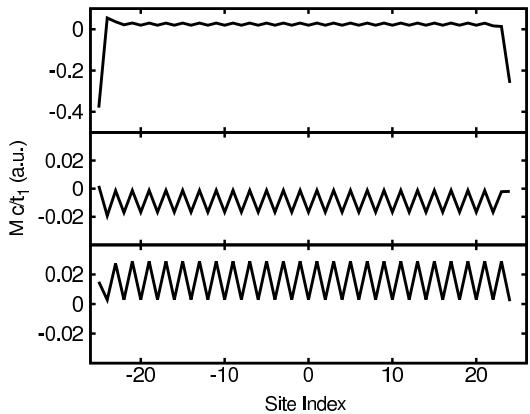}
\caption{Local magnetization for a normal insulator---point (a) in the phase diagram---along the line shown in Fig. \protect\ref{fig:flake}. Top panel: site contributions to the trace in \protect\equ{m3}. Middle panel: first term in \protect\equ{mpq}. Bottom panel: second term in \protect\equ{mpq}.
Notice the different scales.}
\label{fig:m1} \end{figure}

\begin{figure}
\centering
\includegraphics[width=.8\columnwidth]{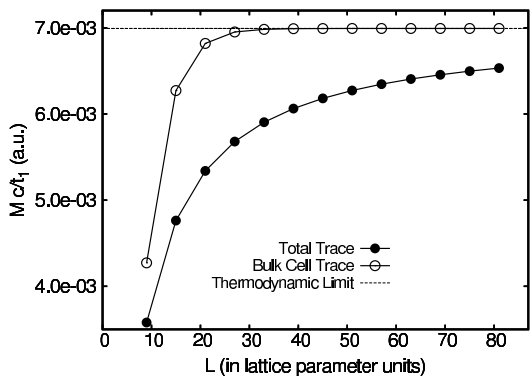}
\caption{Convergence of $M$ with the flake size, point (a) in the phase diagram. Filled circles: total magnetic moment divided by the flake area, \protect\eqs{m3}{local}. Open circles: average of $\MM(i)$ over the two central sites in the flake. All flakes have the same aspect ratio as in Fig. \protect\ref{fig:flake}; the abscissae indicate the lenght of the arm-chair edge in lattice parameter units (75 in Fig. \protect\ref{fig:flake}).
}
\label{fig:m2} \end{figure}

We illustrate the case of a normal insulator ($C=0$), choosing the point (a) in the phase diagram ($\Delta/t_2= 3.67$, $\varphi=0.1 \pi$). We address, within OBCs, finite flakes of rectangular shape cut from the bulk, as shown in Fig. \ref{fig:flake}. We separately plot in the two lowest panels of Fig. \ref{fig:m1} the two terms in \equ{mpq}: they correspond to the ``local circulation'' and ``itinerant circulation'', respectively, in the language of Refs. \cite{rap128,rap130}. It is easily realized that both are bulk, i. e. their average over any bulk cell coincides (in the large-$A$ limit) with the average over the whole sample. The top panel of Fig. \ref{fig:m1} shows, by contrast, the site contributions to \equ{m3}. Although the trace is the same as in \equ{local}, the difference is striking: here most of the magnetization is due to the boundary.
We then show in Fig. \ref{fig:m2} the convergence of the computed $M$ with the flake size. The figure shows that the macroscopic average of $\MM(i)$ in the flake center converges much faster than the trace, \eqs{m3}{local}. The former converges exponentially, owing to the density matrix decay; the latter shows a $1/L$ convergence, because the number of bulk sites scales as $L^2$, while the number of boundary sites scales as $L$.

Next, we address Chern insulators. Therein the spectrum of a finite sample within OBCs becomes gapless in the large sample limit; when $\mu$ is in the bulk gap, the bulk is insulating but $M$ depends on $\mu$, owing to boundary currents. We are going to prove that even this extra contribution to $M$ is bulk in the above sense.

The macroscopic magnetization of a 2D macroscopic sample at fixed chemical potential is  $M = - (1/A)\, \partial G / \partial B$, where $G$ is the Gibbs grand potential. At zero temperature $G = U - \mu N$, and $\mu$ is the Fermi level: \beq M = -\frac{1}{A}\frac{\partial U}{\partial B} + \frac{\mu}{A}\frac{\partial N}{\partial B} = M_1 + M_2  . \label{G} \eeq It is easy to show, using the Hellmann-Feynman theorem, that the first term $M_1$ in \equ{G} coincides with \equ{m2}, hence also with \equ{local}. Defining the areal density $n = N/A$, the second term in \equ{G} is  $M_2 = \mu \, \partial n / \partial B$; we then make contact with St\v{r}eda's formula \cite{Streda82}  \beq \frac{\partial n}{\partial B} = \frac{e C}{2 \pi \hbar c} = \frac{C}{\phi_0} , \label{streda}  \eeq where $C$ is the Chern number and $\phi_0 = hc/e$ is the flux quantum. The formula was proved for a crystalline 2D   system within PBCs \cite{Xiao05,Xiao10}. Its OBCs analogue displays subtle features, since for an isolated sample the number of electrons $N$ stays constant: we are going to show that the boundary acts as a reservoir, in such a way that the density $n$ in the bulk region obeys indeed St\v{r}eda's formula.

Even the Chern number admits a local description in real space \cite{rap146}, and can be directly expressed in terms of the ground state density matrix within either PBCs or OBCs. Here we define the dimensionless function \cite{nota2} \beq \CC(\r) =  4 \pi \,\mbox{Im } \me{\r}{\, \CQ x \CP y \CQ \,}{\r} , \label{chern} \eeq whose macroscopic average in the bulk of a sample equals $C$. Therefore the magnetization of an insulator---either normal or Chern---obtains from the macroscopic average of $\MM(\r) = \MM_1(\r) + \MM_2(\r)$ in some inner region of the sample, where $\MM_1(\r)$ is the same as in \equ{mpq}, and \beq \MM_2(\r) = \frac{\mu}{\phi_0} \CC(\r) = \mu \frac{2e}{\hbar c} \mbox{Im } \me{\r}{\, \CQ x \CP y \CQ \,}{\r} . \eeq We may also rewrite the local magnetization as \bea \MM(\r) &=& \frac{e}{\hbar c} \mbox{Im }  \me{\r}{ \CP x \CQ H \CQ y \CP }{\r} \nn &-& \frac{e}{\hbar c} \mbox{Im } \me{\r}{ \CQ x \CP ( H - 2 \mu ) \CP y \CQ }{\r} . \label{mmu} \eea It is easy to verify that the macroscopic average of $\MM(\r)$ is invariant by translation of the energy zero, i.e. is invariant under the transformation $H \rightarrow H + \Delta \epsilon$, $\mu \rightarrow \mu + \Delta \epsilon$, as it must be. If the thermodynamic limit is taken {\it before} the trace, \equ{mmu} can be related to the known $\k$-space theory for the magnetization of a Chern insulator \cite{rap130}, and also to a recent reformulation and generalization to disordered systems \cite{Schulz12}.

\begin{figure}
\centering
\includegraphics[width=.8\columnwidth]{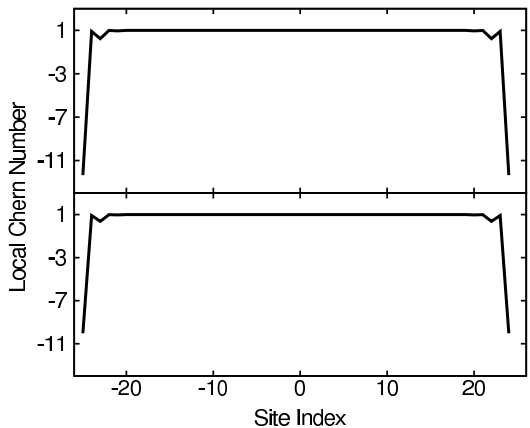}
\caption{Point (b) in the phase diagram.
Top panel: local Chern number, \equ{chern}. Bottom panel: $B$-derivative of the density in dimensionless units, \equ{chern2}.}
\label{fig:chern} \end{figure}

We stress a very crucial feature. In \equ{local} we have taken the trace of $\MM_1(\r)$ over the whole sample within OBCs; we cannot do the same with $\MM_2(\r)$, because such trace---as well as the trace of $\CC(\r)$---identically vanishes. We show in Fig. \ref{fig:chern}, top panel, a plot of $\CC(i)$: for the sake of simplicity we choose the very high symmetry point (b) in the phase diagram ($\Delta=0, \varphi=\pi/2$), where the site occupancy in the bulk region is $n(i) = 1/2$, and $M_1=0$. It is perspicuous that the local Chern numbers $\CC(i)$ are equal to 1 in the bulk of the sample, while they deviate and become negative in the boundary region.

One gets the bulk magnetization $M$ at fixed $\mu$ by taking the macroscopic average of \equ{mmu} in the relevant sample region, for both normal and Chern insulators \cite{note}. In the case of our tight-binding model, it is enough to take the average over the two central sites of the flake. 

In order to evaluate $M$ there is no need of running finite-$B$ calculations; nonetheless it is worth showing how St\v{r}eda's formula works within OBCs. To this aim we use \equ{streda} in reverse: we give an alternative form for the local Chern number as
\beq \tilde{\CC}(\r) = \phi_0 \frac{\partial n(\r)}{\partial B} , \label{chern2} \eeq and we evaluate the $B$-derivative numerically.  The result is shown in Fig. \ref{fig:chern}, bottom panel, for a $B$ value such that the flux through the unit cell is $\phi = 0.001 \phi_0$. The plot of the $B$-derivative of the density, as in \equ{chern2}, shows that St\v{r}eda's formula holds even {\it locally}, and confirms that the boundary region acts as an electron reservoir. 


Our presentation has been limited to the 2D case for the sake of clarity; but the 3D theory is not conceptually different, although it requires a more complex algebra.
In conclusion, we have shown that the orbital magnetization $\M$ in any macroscopically homogeneous region of an insulator---even topologically nontrivial---obtains as the macroscopic average of a magnetization density $\MM(\r)$, \equ{mmu}, uniquely defined in terms of the density matrix in a neighborhood of $\r$, and insensitive to the conditions of the sample boundary. The approach applies with no major changes to disordered materials as well. Polarization $\P$ behaves differently: a polarization density enjoying a similar property cannot be defined.

R. R.  thanks H. Schulz-Baldes for bringing Ref. \cite{Schulz12} to his attention.
Work partially supported by the ONR Grant No. N00014-11-1-0145.


\begin{thebibliography}{10}
\bibitem{nota1}
{ The statement does not apply to spin magnetization, where the ``dipolar
  density'' is trivially proportional to the spin density}.
\bibitem{nota2}
{ With respect to $\CC(\r)$ as defined in Ref. \protect\cite{rap146}, here it
  is expedient to exchange $\CP$ with $\CQ$, and consequently to change sign}.
\end{thebibliography}

\end{document}